\documentclass[titlepage,12pt]{article}
\usepackage{amsmath}
\usepackage{graphicx}
\usepackage{amsfonts}
\usepackage{amssymb}

\input{epsf}
\begin{document}

\title{Nonlocal Equation of State\\in \\Anisotropic Static Fluid Spheres in General Relativity}
\author{\textbf{H. Hern\'{a}ndez}\\{\textit{Laboratorio de F\'{\i}sica Te\'{o}rica, Departamento de F\'{\i}sica,}}\\{\textit{Facultad de Ciencias, Universidad de Los Andes,}}\\{\textit{M\'{e}rida 5101, Venezuela} }\\and \\\textit{Centro Nacional de C\'alculo Cient\'{\i}fico}\\\textit{Universidad de Los Andes (CeCalCULA),}\\\textit{Corporaci\'{o}n Parque Tecnol\'{o}gico de M\'{e}rida,}\\\textit{M\'{e}rida 5101, Venezuela}
\and \textbf{L.A. N\'{u}\~{n}ez}\\\textit{Centro de Astrof\'{\i}sica Te\'{o}rica,} \textit{Facultad de Ciencias,}\\\textit{Universidad de Los Andes,} \textit{M\'{e}rida 5101, Venezuela}\\and \\\textit{Centro Nacional de C\'alculo Cient\'{\i}fico}\\\textit{Universidad de Los Andes (CeCalCULA),}\\\textit{Corporaci\'{o}n Parque Tecnol\'{o}gico de M\'{e}rida, }\\\textit{M\'{e}rida 5101, Venezuela}}
\date{March 2002}
\maketitle
\begin{abstract}
We show that it is possible to obtain credible static anisotropic spherically
symmetric matter configurations starting from known density profiles and
satisfying a nonlocal equation of state. These particular types of equation of
state describe, at a given point, the components of the corresponding
energy-momentum tensor not only as a function at that point, but as a
functional throughout the enclosed configuration. To establish the physical
plausibility of the proposed family of solutions satisfying nonlocal equation
of state, we study the constraints imposed by the junction and energy
conditions on these bounded matter distributions.

We also show that it is possible to obtain physically plausible static
anisotropic spherically symmetric matter configurations, having nonlocal
equations of state\textit{, }concerning the particular cases where the radial
pressure vanishes and, other where the tangential pressures vanishes. The
later very particular type of relativistic sphere with vanishing tangential
stresses is inspired by some of the models proposed to describe extremely
magnetized neutron stars (magnetars) during the transverse quantum collapse.
\end{abstract}

\section{Introduction}

Classical continuum theories are based on the assumption that the state of a
body is determined entirely by the behavior of an arbitrary infinitesimal
neighborhood centered at any of its material points. Furthermore, there is
also a premise that any small piece of the material can serve as a
representative of the entire body in its behavior, hence the governing balance
laws are assumed to be valid for every part of the body, no matter how small.
Clearly, the influence of the neighborhood on motions of the material points,
emerging as a result of the interatomic interaction of the rest of the body,
is neglected. Moreover, the isolation of an arbitrary small part of the body
to represent the whole clearly ignores the effects of the action of the
applied load at a distance. These applied loads are important because their
transmissions from one part of the body to another, through their common
boundaries affect the motions, hence the state of the body at every point. The
relevance of long-range or nonlocal outcomes on the mechanical properties of
materials are well known. The main ideas of Nonlocal Continuum Mechanics were
introduced during the 1960s (see \cite{Narasimhan1993} and references therein)
and are based on considering the stress to be a function of the mean of the
strain from a certain representative volume of the material centered at that
point. Since then, there have been many situations of common occurrence
wherein nonlocal effects seem to dominate the macroscopic behavior of matter.
Interesting problems coming from a wide variety of areas such as damage and
cracking analysis of materials, surface phenomena between two liquids or two
phases, mechanics of liquid crystals, blood flow, dynamics of colloidal
suspensions seem to demand this type of nonlocal approach which has made this
area very active concerning recent developments in material and fluid science
and engineering.

The true equation of state that describes the properties of matter at
densities higher than nuclear ($\approx10^{14}$ $gr/cm^{3}$) is essentially
unknown due to our inability to verify the microphysics of nuclear matter at
such high densities \cite{Demianski1985}-\cite{Glendening2000}. Currently,
what is known in this active field comes from the experimental insight and
extrapolations from the ultra high energy accelerators and cosmic physics (see
\cite{HeiselbergPandharipande2000} and references therein). Having this
uncertainty in mind, it seems reasonable to explore what is allowed by the
laws of physics, in particular considering spherical or axial symmetries
within the framework of the theory of General Relativity. In the present
paper, we shall consider a \textit{Nonlocal Equation of State} (\textit{NLES}
from now on) in general relativistic static spheres. It follows a previous
work \cite{HernandezNunezPercoco1998} on collapsing configurations having a
\textit{NLES.} The static limit of the particular \textit{NLES} considered in
this reference can be written as
\begin{equation}
P_{r}(r)=\rho(r)-\frac{2}{r^{3}}\int_{0}^{r}\bar{r}^{2}\rho(\bar
{r})\ \mathrm{d}\bar{r}\ +\frac{\mathcal{C}}{2\pi r^{3}}\ ; \label{globalst}%
\end{equation}
where $\mathcal{C}$ is an arbitrary integration constant. It is clear that in
equation (\ref{globalst}) a collective behavior on the physical variables
$\rho(r)$ and $P_{r}(r)$ is present. The radial pressure $P_{r}(r)$ is not
only a function of the energy density, $\rho(r),$ at that point but also its
functional throughout the rest of the configuration. Any change in the radial
pressure takes into account the effects of the variations of the energy
density within the entire volume.

Additional physical insight of the nonlocality for this particular equation of
state can be gained by considering equation (\ref{globalst}) re-written as
\begin{equation}
P_{r}(r)=\rho(r)-\frac{2}{3}\left\langle \rho\right\rangle _{r}+\frac
{\mathcal{C}}{r^{3}}\ ,\qquad\mathrm{with\quad}\left\langle \rho\right\rangle
_{r}=\frac{\int_{0}^{r}4\pi\bar{r}^{2}\rho(\bar{r})\ \mathrm{d}\bar{r}}%
{\frac{4\pi}{3}r^{3}}\ =\frac{M(r)}{V(r)}\ . \label{averagedens1}%
\end{equation}
Clearly the nonlocal term represents an average of the function $\rho(r)$ over
the volume enclosed by the radius $r$. Moreover, equation (\ref{averagedens1})
can be easily rearranged as
\begin{equation}
P_{r}(r)=\frac{1}{3}\rho(r)+\frac{2}{3}\ \left(  \rho(r)-\left\langle
\rho(r)\right\rangle \right)  \ +\frac{\mathcal{C}}{r^{3}}=\frac{1}{3}%
\rho(r)+\frac{2}{3}\ \mathbf{\sigma}_{\rho}+\frac{\mathcal{C}}{r^{3}}\ ,
\label{sigmatmunu}%
\end{equation}
where we have used the concept of statistical standard deviation
$\mathbf{\sigma}_{\rho}$ from the local value of energy density. Furthermore,
we may write:%

\begin{equation}
P_{r}(r)=\mathcal{P}(r)+2\mathbf{\sigma}_{\mathcal{P}(r)}+\frac{\mathcal{C}%
}{r^{3}}\ \quad\mathrm{where\quad}\left\{
\begin{array}
[c]{l}%
\mathcal{P}(r)=\frac{1}{3}\rho(r)\qquad\mathrm{and}\\
\\
\mathbf{\sigma}_{\mathcal{P}(r)}=\left(  \frac{1}{3}\rho(r)-\frac{1}%
{3}\left\langle \rho\right\rangle _{r}\right)  =\left(  \mathcal{P}%
(r)-{\bar{\mathcal{P}}}(r)\right)  \ .
\end{array}
\right.  \label{sigma0}%
\end{equation}
Therefore, if at a particular point within the distribution the value of the
density, $\rho(r),$ gets very close to its average $\left\langle
\rho(r)\right\rangle $ the equation of state of the material becomes similar
to the typical radiation dominated environment, $P_{r}(r)\approx
\mathcal{P}(r)\equiv\frac{1}{3}\rho(r).$

In reference \cite{HernandezNunezPercoco1998}, it is shown that under
particular circumstances a general relativistic spherically symmetric
anisotropic (nonequal radial and tangential pressures, i.e. $P_{r}%
~\neq~P_{\perp}$) distribution of matter could satisfy a \textit{NLES}. Some
of these dynamic bounded matter configurations having a \textit{NLES} with
constant gravitational potentials at the surface, admit a Conformal Killing
Vector and fulfill the energy conditions for anisotropic imperfect fluids.
Several analytical and numerical models for collapsing radiating anisotropic
spheres in general relativity were also developed in that paper. Although the
perfect pascalian fluid assumption (i.e. $P_{r}~=~P_{\perp}$) is supported by
solid observational and theoretical grounds, an increasing amount of
theoretical evidence strongly suggests that a variety of very interesting
physical phenomena may take place giving rise to local anisotropy. In the
Newtonian regime, the consequences of local anisotropy originated by
anisotropic velocity distributions have been pointed out in the classical
paper by J.H. Jeans \cite{Jeans1922}. In the context of General Relativity, it
was early remarked by G. Lema\^{i}tre \cite{Lemaitre1933} that local
anisotropy can relax the upper limits imposed on the maximum value of the
surface gravitational potential. Since the pioneering work of R. Bowers and E.
Liang \cite{BowerLiang1974}, the influence of local anisotropy in General
Relativity has been extensively studied (see \cite{HerreraSantos1997} and
references therein). Recently, the role of local anisotropy in the equation of
state modeling the interior of extremely magnetized neutron stars or Magnetars
at supranuclear densities, has been attracting the attention of researchers
(see \cite{SaengMathews1999}-\cite{KohriYamadaNagataki2001} and references
therein). It has been shown that these intense magnetic fields could induce
local anisotropy in pressures within the equation of state of the matter
distribution. In fact, two of these references\cite{ChaichianEtal1999} and
\cite{PerezMartinezEtal2000} consider the consequences on the instability of
the matter configuration in the most extreme case of anisotropy in pressures,
i.e. when there exists a vanishing pressure perpendicular to the magnetic
field. In this case the vanishing of the pressure transverse to the magnetic
field leads the configuration to a transverse quantum collapse of the configuration.

The present paper is focused on the plausibility of obtaining static solutions
satisfying \textit{NLES} and whether those solutions could represent a
credible bounded anisotropic matter distributions in General Relativity. We
shall show that it is possible to obtain static anisotropic spherically
symmetric matter configurations starting from known density profiles and
satisfying a \textit{NLES.} We also present two static solutions for
\textit{NLES} configurations concerning the particular case of vanishing
radial pressure ($P_{\perp}\neq0$ and $P_{r}=0$) and vanishing tangential
stresses, i.e., $P_{\perp}=0$ and $P_{r}\neq0.$

The structure of the present work is the following. The next section contains
an outline of the general conventions, notation used, the metric and the
corresponding field equations. Section \ref{Einsteinfield} is devoted to solve
the Einstein Field Equations for a matter distribution assuming a nonlocal
equation of state. The family of \textit{NLES} is presented in Section
\ref{family}. In Section \ref{junction} we study the consequences imposed by
the junction and energy conditions on bounded matter distribution. The method
and some fluid sphere models are considered in the Section \ref{themethod}.
The two particular cases, i.e. $P_{r}=0,$ and $P_{\perp}=0,$ are presented in
Section \ref{magnetar}. Finally, in the last section, our concluding remarks
and results are summarized.

\section{The Einstein Field Equations}

\label{Einsteinfield}To explore the feasibility of nonlocal equations of state
for bounded configurations in General Relativity, we shall consider a static
spherically symmetric anisotropic distribution of matter with an
energy-momentum represented by $\mathbf{T}_{\nu}^{\mu}~=~{diag}\,~(\rho
,-P_{r},-P_{\perp},-P_{\perp})$, where, $\rho$ is the energy density, $P_{r}$
the radial pressure and $P_{\perp}$ the tangential pressure.

We adopt standard Schwarzschild coordinates $(t,r,\theta,\phi)$ where the line
element can be written as
\begin{equation}
\mathrm{d}s^{2}=e^{2\nu(r)}\mathrm{d}t^{2}-e^{2\lambda(r)}\mathrm{d}%
r^{2}-r^{2}\mathrm{d}\Omega^{2}\,, \label{metrica}%
\end{equation}
with $d\Omega^{2}\equiv d\theta^{2}+\sin^{2}\theta d\phi^{2}$ , the solid angle.

The resulting Einstein equations are:
\begin{align}
8\pi\rho &  =\frac{1}{r^{2}}+\frac{e^{-2\lambda}}{r}\left[  2\lambda^{\prime
}-\frac{1}{r}\right]  \,,\label{ee1}\\
& \nonumber\\
-8\pi P_{r}  &  =\frac{1}{r^{2}}-\frac{e^{-2\lambda}}{r}\left[  2\nu^{\prime
}+\frac{1}{r}\right]  \,\quad\qquad\mathrm{and}\label{ee2}\\
& \nonumber\\
-8\pi P_{\perp}  &  =e^{-2\lambda}\left[  \frac{\lambda^{\prime}}{r}-\frac
{\nu^{\prime}}{r}-\nu^{\prime\prime}+\nu^{\prime}\lambda^{\prime}-(\nu
^{\prime})^{2}\right]  \,, \label{ee3}%
\end{align}
where primes denote differentiation with respect to $r$.

Using equations (\ref{ee2}) and (\ref{ee3}), or equivalently the conservation
law ${\mathbf{T}_{\nu}^{\mu}}_{;\mu}~=~0$, we obtain the hydrostatic
equilibrium equation for anisotropic fluids
\begin{equation}
P_{r}^{\prime}=-\left(  \rho+P_{r}\right)  \nu^{\prime}+\frac{2}{r}\left(
P_{\perp}-P_{r}\right)  \,. \label{eeh}%
\end{equation}

Equation (\ref{ee1}) can be formally integrated to give
\begin{equation}
e^{-2\lambda}=1-2\frac{m(r)}{r}\,, \label{explam}%
\end{equation}
where a \textit{mass function} $m(r),$ has been defined by
\begin{equation}
m(r)=4\pi\,\int_{0}^{r}\rho\,{{\bar{r}}^{2}\,\mathrm{d}{\bar{r}}}\,,
\label{eme}%
\end{equation}
and it corresponds to the mass inside a sphere of radius $r$ as seen by a
distant observer.

Finally, from (\ref{eeh}), (\ref{explam}) and (\ref{ee2}) the anisotropic
Tolman-Oppenheimer-Volkov (TOV) equation \cite{BowerLiang1974} can be written
as
\begin{equation}
\frac{\mathrm{d\,}P_{r}}{\mathrm{d}\,r}=-\left(  \rho+P_{r}\right)  \left(
\frac{m+4\pi r^{3}P_{r}}{r\left(  r-2m\right)  }\right)  +\frac{2}{r}\left(
P_{\perp}-P_{r}\right)  \,. \label{anitov}%
\end{equation}
Obviously, in the isotropic case $(P_{\perp}=P_{r})$ it becomes the usual TOV equation.

It has been established \cite{RendallSchmidt91}-\cite{MarcMerceSenovilla2002}
that if $\rho$ is a continuous positive function, $P_{\perp}(r)$ is a
continuous differentiable function and $P_{r}(r)$ is a solution to the
equation with starting value $P_{\perp}(0)=P_{r}(0),$ there exists a unique
global solution to (\ref{anitov}) representing a spherically symmetric fluid
ball in General Relativity.

\section{A Family of Solutions with a \textit{NLES}}

\label{family}In this section we are going to present a family of static
solution of the Einstein equations satisfying a \textit{NLES. }Defining the
new variables:
\begin{equation}
e^{2\nu\left(  r\right)  }=h\left(  r\right)  \,e^{4\beta\left(  r\right)
},\quad\mathrm{and}\qquad e^{2\lambda\left(  r\right)  }=\frac{1}{h\left(
r\right)  };\quad\mathrm{\qquad with}\qquad h(r)\equiv1-2\frac{m\left(
r\right)  }{r}\ ,\label{elemetric}%
\end{equation}
the above metric (\ref{metrica}) can be re-written as%

\begin{equation}
\mathrm{d}s^{2}=h\left(  r\right)  \,e^{4\beta(r)}\mathrm{d}t^{2}-\frac
{1}{h(r)}dr^{2}-r^{2}\mathrm{d}\Omega^{2}\,, \label{metrica2}%
\end{equation}
The resulting Einstein Equations are:
\begin{align}
8\pi\rho &  =\frac{1-h-h^{\prime}r}{r^{2}}\,,\label{eeg1}\\
& \nonumber\\
8\pi P_{r}  &  =-\frac{1-h-h^{\prime}r}{r^{2}}+\frac{4\,h\,\beta^{\prime}}%
{r}\qquad\mathrm{and}\label{eeg2}\\
& \nonumber\\
8\pi P_{\perp}  &  =\frac{\,h^{\prime}+2\,h\,\beta^{\prime}}{r}+\frac{1}%
{2}\left[  h^{\prime\prime}+4\,h\,\beta^{\prime\prime}+6\,h^{\prime}%
\beta^{\prime}+8\,h\,\left(  \beta^{\prime}\right)  ^{2}\right]  \,.
\label{eeg3}%
\end{align}
Now, equation (\ref{globalst}) can re-stated as a differential equation%
\begin{equation}
P_{r}(r)=\rho(r)-\frac{2}{r^{3}}\int_{0}^{r}\bar{r}^{2}\rho(\bar
{r})\ \mathrm{d}\bar{r}\ +\frac{\mathcal{C}}{2\pi r^{3}}\qquad\Leftrightarrow
\qquad\rho-3P_{r}+r\left(  \rho^{\prime}-P_{r}^{\prime}\right)  =0\ .
\label{eeg}%
\end{equation}
Thus, from (\ref{eeg}), (\ref{eeg1}) and (\ref{eeg2}), we have
\begin{equation}
\frac{2}{r}\left(  h^{\prime}+2h\beta^{\prime}\right)  +h^{\prime\prime
}+2\beta^{\prime}h^{\prime}+2h\beta^{\prime\prime}=0\ ,
\end{equation}
which can be formally integrated yielding
\begin{equation}
\beta(r)=\frac{1}{2}\ln\left(  \frac{\mathrm{C}}{h}\right)  +\int
\frac{\mathcal{C}}{r^{2}\,h}\ \mathrm{d}\,r+C_{1}\,, \label{beta}%
\end{equation}
where $\mathrm{C}$ and $C_{1}$ are arbitrary integration constants.

At this point, equation (\ref{beta}) deserves several comments:

\begin{itemize}
\item  Firstly, if we set ${\mathcal{C}}~=~C_{1}~=~0$ , a particular family of
solutions ,
\begin{equation}
\beta(r)=\frac{1}{2}\ln\left(  \frac{\mathrm{C}}{h}\right)  ,\label{oldsoluc}%
\end{equation}
considered in a previous work \cite{HernandezNunezPercoco1998} is found. The
metric obtained in this particular static case recalls the so called
isothermal coordinates system \cite{Synge1960} which in turn is a particular
case of the more general ``warped space-time'' (we refer the reader to
\cite{CarotDaCosta93} and references cited therein for a general discussion)

\item  Next, is the approach we have followed in order to obtain static
anisotropic solutions having a \textit{NLES}. It is clear that if the profile
of the energy density, $\rho(r)$, is provided, the metric elements $h(r)$ and
$\beta(r)$ can be calculated through (\ref{eme}), (\ref{elemetric}) and
(\ref{beta}). Therefore, we can develop a consistent method to obtain static
solutions having \textit{NLES} from a known static ones.

\item  Finally, the metric elements (\ref{elemetric}) and (\ref{beta}%
),\ should fulfill the junction conditions and the physical variables coming
from the energy momentum tensor, are only restricted by the hydrostatic
equilibrium equation (\ref{anitov}) and by some elementary criteria of
physical acceptability. The next section is devoted to list these criteria for
anisotropic and isotropic fluids with a \textit{NLES}.
\end{itemize}

In terms of the metric elements (\ref{elemetric}) and \ref{beta}, the
corresponding Einstein Field Equations for anisotropic fluids having
\textit{NLES }can be written as:
\begin{align}
8\pi\rho &  =\frac{2m^{\prime}}{r^{2}},\label{rho}\\
& \nonumber\\
8\pi P_{r}  &  =\frac{2m^{\prime}}{r^{2}}-\frac{4\left(  m-\mathcal{C}\right)
}{r^{3}},\qquad\qquad\mathrm{and}\label{Prad}\\
& \nonumber\\
8\pi P_{\perp}  &  =\frac{m^{\prime\prime}}{r}+\frac{2\left(  m^{\prime
}r-m\right)  }{r^{3}}\left[  \frac{m^{\prime}r-m}{r-2m}-1\right]  .
\label{Ptang}%
\end{align}

\section{Junction and Energy Conditions}

\label{junction}Most exact solutions\ of the differential equation
(\ref{anitov}) supplied by the literature have been obtained from excessively
simplifying assumptions solely with the purpose to find such solutions, and,
consequently, they rarely represent physically ``realistic'' fluids (see for
example, two interesting and complementary reviews on this subject:
\cite{FinchSkea} and \cite{DelgatyLake1998}). In order to establish the
physical acceptability of the proposed static family of solutions
(\ref{elemetric}) and (\ref{beta}), satisfying a \textit{NLES,}
(\ref{globalst}) we state the consequences imposed by the junction and energy
conditions for anisotropic fluids on bounded matter distribution.

We are going to consider bounded configurations, i.e. a matter distribution
isolated in the sense that, where the pressure vanishes occurs at a finite
radius. In fact, it can easily be shown that the necessary and sufficient
condition for matching an interior solution (\ref{metrica2}) onto the exterior
Schwarzschild solution of total mass $M$,
\begin{equation}
\mathrm{d}s^{2}=\left(  1-2\frac{M}{r}\right)  \mathrm{d}t^{2}-\left(
1-2\frac{M}{r}\right)  ^{-1}\mathrm{d}r^{2}-r^{2}\mathrm{d}\Omega^{2}\,,
\end{equation}
is that the pressure equals zero at a finite radius $r=a$. In this case, the
interior metric (\ref{metrica2}) should satisfy the following conditions at
the boundary surface of the sphere $r=a:$
\begin{equation}
P_{r}(a)=0\qquad\mathrm{\Rightarrow}\qquad\beta\left(  a\right)  =\beta
_{a}=0,\qquad\mathrm{and}\qquad m(a)=M\ . \label{cda}%
\end{equation}
>From the condition $P_{r}(a)=0,$ it is clear that
\begin{equation}
M=4\pi\,\int_{0}^{a}\rho\,{{\bar{r}}^{2}\,\mathrm{d}{\bar{r}}}\,=2\pi
a^{3}\rho\left(  a\right)  \,. \label{constantC}%
\end{equation}
Now, equation (\ref{Prad}) leads to ${\mathcal{C}}=0$ i.e.%
\begin{equation}
m(r)={\mathcal{C}}+2\pi\left[  \rho(r)-P_{r}(r)\right]  r^{3}\qquad
\mathrm{if}\quad m(0)=0\qquad\mathrm{\Rightarrow}\quad{\mathcal{C}}=0\ .
\label{zeroc}%
\end{equation}

We should require that the metric elements have to be finite and non zero
everywhere within the matter configuration, with no changes of sign nor loss
of reality allowed. This implies that
\begin{equation}
m(r)>\frac{r}{2}\qquad\qquad\forall\ r\,,\label{mas2r}%
\end{equation}
where $m(r)$ is the mass function defined by equation (\ref{eme}). Recently
\cite{MarcMerceSenovilla2002} the above condition (\ref{mas2r}), for
spherically static bounded configurations, has been proved to be equivalent to
requiring that $\rho~+~P_{r}~+~2~\,P_{\perp}~\geq~0,$ (i.e. the \textit{strong
energy condition} ) which clearly implies that the space time cannot contain a
black hole region.

Next, we will establish the consequences of the physical acceptability for an
anisotropic fluid configuration in terms of the mass function $m(r)$.

\begin{enumerate}
\item  The density must be positive definite and its gradient must be negative
everywhere within the matter distribution. Trivially, from equation
(\ref{rho}) and its derivative we obtain
\begin{equation}
\rho>0\qquad\mathrm{\Rightarrow}\qquad m^{\prime}>0\qquad\mathrm{and}%
\qquad\frac{\partial\rho}{\partial r}<0\qquad\mathrm{\Rightarrow}\qquad
m^{\prime\prime}<\frac{2m^{\prime}}{r}\ . \label{rhopost}%
\end{equation}

\item  The radial and tangential pressure must be positive definite.
Therefore, equations (\ref{Prad}) and (\ref{Ptang}), yield
\begin{align}
P_{r}  &  \geq0\qquad\mathrm{\Rightarrow}\qquad m^{\prime}\geq\frac{2m}%
{r}\qquad\mathrm{and}\label{presrposit}\\
& \nonumber\\
P_{\perp}  &  \geq0\qquad\mathrm{\Rightarrow}\qquad m^{\prime\prime}\geq
-\frac{2\left(  m^{\prime}r-m\right)  }{r^{2}}\left[  \frac{m^{\prime}%
r-m}{r-2m}-1\right]  \ . \label{presTGposit}%
\end{align}

\item  The radial pressure gradient must be negative definite. Thus,
differentiating equation (\ref{Prad}) we get
\begin{equation}
\frac{\partial P_{r}}{\partial r}\leq0\qquad\mathrm{\Rightarrow}\qquad
m^{\prime\prime}\leq\frac{4m^{\prime}}{r}-\frac{6m}{r^{2}}\ . \label{Dpres}%
\end{equation}

\item  The speed of sound should be subluminal, consequently from equation
(\ref{eeg}) we obtain
\begin{equation}
{v_{s}}^{2}\equiv\frac{\partial P_{r}}{\partial\rho}\leq1\,\qquad
\mathrm{\Rightarrow}\qquad\frac{\left(  \rho-3\,P_{r}\,\right)  }{r\left(
\frac{\partial\rho}{\partial r}\right)  }\leq0\ . \label{velsonid}%
\end{equation}
It is clear that, due to the density gradient being negative everywhere within
the configuration, the requirement of subluminal sound speeds leads to
$\rho>3\,P_{r}\,$for both anisotropic and isotropic fluids having a
\textit{NLES}.
\end{enumerate}

In addition to the above intuitive conditions we should satisfy either the
\textit{Strong Energy Condition} or \textit{Dominant Energy Condition:}

\begin{itemize}
\item  Demanding that the trace of the energy-momentum tensor be positive we
find for the \textit{Strong Energy Condition}, $\rho+P_{r}+2\,P_{\perp}\geq0$.
Thus, using equations (\ref{rho}), (\ref{Prad}) and (\ref{Ptang}) we get
\begin{equation}
\rho+P_{r}+2\,P_{\perp}\geq0\qquad\mathrm{\Rightarrow\ }m^{\prime\prime}%
\geq-\frac{2m^{\prime}}{r}+\frac{2m}{r^{2}}-\frac{2\left(  m^{\prime
}r-m\right)  }{r^{2}}\left[  \frac{m^{\prime}r-m}{r-2m}-1\right]
.\label{strgenerani}%
\end{equation}
The Strong Energy Condition also implies $\rho+P_{r}\geq0$ and $\rho+P_{\perp
}\geq0$, therefore,
\begin{align}
\rho+P_{r} &  \geq0\qquad\mathrm{\Rightarrow}\qquad m^{\prime}\geq\frac{m}%
{r}\qquad\mathrm{and}\label{strgenerani1}\\
& \nonumber\\
\rho+P_{\perp} &  \geq0\qquad\mathrm{\Rightarrow}\qquad m^{\prime\prime}%
\geq-\frac{2m^{\prime}}{r}-\frac{2\left(  m^{\prime}r-m\right)  }{r^{2}%
}\left[  \frac{m^{\prime}r-m}{r-2m}-1\right]  \ .\label{strgenerani2}%
\end{align}

In the case of isotropic fluids, the above condition (\ref{strgenerani})
reads
\begin{equation}
\rho\geq3\,P_{r}\qquad\mathrm{\Rightarrow}\qquad m^{\prime}\leq\frac{3m}{r}\,.
\label{strgeneriso}%
\end{equation}

\item  The \textit{Dominant Energy Condition} entails that the density must be
larger than the pressure. Now, by using equations (\ref{rho}), (\ref{Prad})
and (\ref{Ptang}) we get
\begin{align}
\rho &  \geq P_{r}\qquad\mathrm{\Rightarrow}\qquad m\geq0\qquad\mathrm{and}%
\label{domener}\\
& \nonumber\\
\rho &  \geq P_{\perp}\qquad\mathrm{\Rightarrow}\qquad m^{\prime\prime}%
\leq\frac{2m^{\prime}}{r}-\frac{2\left(  m^{\prime}r-m\right)  }{r^{2}}\left[
\frac{m^{\prime}r-m}{r-2m}-1\right]  \ . \label{domenetg}%
\end{align}
\end{itemize}

Notice that, if the density, the radial and the tangential pressures are
positive definite and the Strong Energy Condition is satisfied, then the
Dominant Energy Conditions (\ref{domener}) and (\ref{domenetg}) are
automatically fulfilled, but the inverse is not true.

In order to determine the physical reasonableness\ of the anisotropic
configurations having a \textit{NLES} we shall explore two different sets of
conditions. Both sets include conditions (\ref{mas2r}) through(\ref{velsonid})
but differ in the selection of the Strong Energy Condition or the Dominant
Energy Condition. In the first group, we shall use the Strong Energy Condition
(\ref{strgenerani}), and because we require subluminal sound velocities
(\ref{velsonid}), $\rho\geq3\,P_{r}$ will be required. Obviously, when
isotropic fluids are considered, the Strong Energy Condition reads $\rho
\geq3\,P_{r}$. and the sound speed will be subluminal in all cases. The second
set again comprises conditions (\ref{mas2r}) through (\ref{velsonid}), the
fulfillment of the Dominant Energy Conditions (equations (\ref{domener}) and
(\ref{domenetg})) and the requirement of $\rho\geq3\,P_{r}$.

\section{A Method for \textit{NLES} Static Solutions}

\label{themethod}In the present section we shall state a general method to
obtain \textit{NLES} static anisotropic spherically symmetric solutions from
known density profiles.

\subsection{A Method for \textit{NLES }Anisotropic Solutions}

Concerning anisotropic fluids, i.e. non pascalian fluids where $P_{r}\neq
P_{\perp},$ the method to obtain \textit{NLES} static spherically symmetric
solutions is as follows:

\begin{enumerate}
\item  Select a static density profile $\rho(r)$, from a known static
solution. Then, the mass distribution function, $m=m(r),$ can be obtained
through equation (\ref{eme}). The junction condition (\ref{cda}) implies the
continuity of $h(a)$ and the expression for the total mass $m(a)=M$ can be procured.

\item  Next, check out where and under what circumstances, all the above
physical and energy conditions, written in terms of the mass function,
$m=m(r),$ are fulfilled. In other words, the mass distribution function
obtained from the density profile selected should satisfy the inequalities
(\ref{mas2r}) thought (\ref{strgenerani1}) (or (\ref{mas2r}) thought
(\ref{velsonid}) and (\ref{domener}) and (\ref{domenetg})), at least for some
region $\left[  r_{1},r_{2}\right]  $ within the matter distribution with
$0\leq r_{1}\leq r_{2}\leq a,$ and for particular values of the physical
parameters that characterize the configuration.

\item  Following, the other metric coefficient, $\beta(r),$ can be found by
using equation (\ref{beta}). Notice that because the boundary conditions
(\ref{zeroc}) the actual expression for $\beta(r)$ will be (\ref{oldsoluc}).

\item  Finally, Einstein Field Equations (\ref{Prad}) through (\ref{Ptang})
provide the expressions for the radial and tangential pressures, $P_{r}$ and
$P_{\perp}$, respectively.

\item  The integration constants are obtained as consequences of the junction
conditions at the boundary surface, $r=a$, i.e. $\beta(a)=0,$ $m(a)=M$ and
$P_{r}(a)=0.$
\end{enumerate}

In order to illustrate the above procedure we shall work out several examples
for six static density profiles borrowed from the literature.

\subsection{Examples of \textit{NLES} Static Solutions}

\label{examples}

\paragraph{Example 1:}

The first example comes from a density profile proposed by B. W. Stewart
\cite{Stewart1982}, to describe anisotropic conformally flat static bounded
configurations:%
\begin{equation}
\rho=\frac{1}{8\,\pi r^{2}}\frac{\left(  e^{2Kr}-1\right)  \left(
e^{4Kr}+8Kre^{2Kr}-1\right)  }{\left(  e^{2Kr}+1\right)  ^{3}}\qquad
\Leftrightarrow\qquad m=\frac{r}{2}\left(  \frac{e^{2Kr}-1}{e^{2Kr}+1}\right)
^{2}\ , \label{Examp1}%
\end{equation}
with $K=const.$

Thus, radial and tangential pressures can be written as
\begin{equation}
P_{r}=\frac{1}{8\pi r^{2}}\frac{\left(  1-e^{2Kr}\right)  \left(
e^{4kr}-8Kre^{2Kr}-1\right)  }{\left(  1+e^{2Kr}\right)  ^{3}}\qquad
\mathrm{and}\qquad P_{\bot}=\frac{2K^{2}e^{4Kr}}{\pi\left[  1+e^{2Kr}\right]
^{4}}\ . \label{radtangpresEx1}%
\end{equation}
The constant $K$ can be found from the boundary condition $M=m(r=a)$ as
\begin{equation}
K=\frac{1}{2a}\ln\left[  \frac{1+\left(  \frac{2M}{a}\right)  ^{\frac{1}{2}}%
}{1-\left(  \frac{2M}{a}\right)  ^{\frac{1}{2}}}\right]  \ ,
\end{equation}
while because the pressure vanishes at the surface $P_{r}(r=a)=0$ we have:
\begin{equation}
e^{4Ka}-8Kae^{2Ka}-1=0\ .
\end{equation}

\paragraph{Example 2:}

The density profile of the second example is found due to P.S. Florides
\cite{Florides1974}, but also, corresponding to different solutions, by
Stewart \cite{Stewart1982} and\ more recently by M. K. Gokhroo and A. L. Mehra
\cite{GokhrooMehra1994}. The Gokhroo-Mehra solution, represents densities and
pressures which, under particular circumstances \cite{Martinez1996}, give rise
to an equation of state similar to the Bethe-B\"{o}rner-Sato newtonian
equation of state for nuclear matter
\cite{Demianski1985,ShapiroTeukolsky1983,BetheBornerSato1970}.%
\begin{equation}
\rho=\frac{\sigma}{8\pi}\left[  1-K\frac{r^{2}}{a^{2}}\right]  \qquad
\Leftrightarrow\qquad m={\frac{\sigma\,{r}^{3}}{6}}\,\left[  1\ -\frac
{3\,K}{5}\frac{{r}^{2}}{{a}^{2}}\right]  \ ,\label{Examp2}%
\end{equation}
with $\sigma$ and $K=const.$

The radial and tangential pressures are obtained from Einstein Field Equations
(\ref{Prad}) through (\ref{Ptang}) and can be expressed as
\begin{equation}
P_{r}=\frac{\sigma}{120\pi a^{2}}\left[  5a^{2}-9Kr^{2}\right]  \qquad
\mathrm{and}%
\end{equation}%
\begin{equation}
P_{\bot}=\frac{\sigma}{120\pi a^{2}}\frac{18\sigma K^{2}r^{6}-5a^{2}%
r^{2}K\left(  3r^{2}+54\right)  +25a^{4}\left(  \sigma r^{2}+3\right)
}{5a^{2}\left(  3-\sigma r^{2}\right)  +3\sigma Kr^{4}}\ ,
\end{equation}

where $\sigma=0.631515$.

The boundary conditions at the surface $r=a$, i.e. $M~=~m(a)$ and
$P_{r}(a)~=~0$, lead to
\begin{equation}
K=\frac{5}{9}\qquad\mathrm{and}\qquad M=\frac{\sigma a^{3}}{9}\ .
\end{equation}

\paragraph{Example 3:}

This solution was discovered by H.B. Buchdahl \cite{Buchdahl1959} and
rediscovered later by \cite{DurgapalBannerji1983}. The corresponding density
profile can be written as.%
\begin{equation}
\rho=\frac{3C}{16\pi}\,{\frac{3+C{r}^{2}}{\left(  1+C{r}^{2}\right)  ^{2}}%
}\qquad\Leftrightarrow\qquad m=\frac{3C}{4}\,{\frac{{r}^{3}}{1+C{r}^{2}}}\ ,
\label{Examp3}%
\end{equation}
with $C=const.$

The method applied to the above density profile leads to
\begin{equation}
P_{r}=\frac{3C}{16\pi}\frac{1-Cr^{2}}{\left(  1+Cr^{2}\right)  ^{2}}%
\qquad\mathrm{and}\qquad P_{\bot}=\frac{3C}{16\pi}\,{\frac{2-C{r}^{2}%
+3\,{C}^{2}{r}^{4}}{\left(  2-C{r}^{2}\right)  \left(  1+C{r}^{2}\right)
^{3}}}\ ,
\end{equation}
where $C=\frac{1}{a^{2}}$ coming from $P_{r}(a)=0$

\paragraph{Example 4}

The matter distribution sketched in this example is borrowed from Tolman IV
isotropic static solution which was originally presented by R.C. Tolman in
1939 \cite{Tolman1939}. Tolman IV static solution is, in some aspects, similar
to the equation of state for a Fermi gas in cases of intermediate central
densities. This same profile (and Tolman IV solution)\ is also found as a
particular case of a more general family of solution in \cite{Korkina1981} and \cite{Durgapal1982}%

\begin{equation}
\rho=\frac{C}{8\,\pi}\left[  \frac{1-3K-3Kx}{\left(  1+2\,x\right)  }%
+\frac{2\left(  1+Kx\right)  }{\left(  1+2\,x\right)  ^{2}}\right]
\qquad\Leftrightarrow\qquad m=-\frac{x^{\frac{3}{2}}}{2C^{\frac{1}{2}}}%
\frac{K\left(  1+x\right)  -1}{\left(  1+2x\right)  }\ . \label{Examp4}%
\end{equation}

Einstein Field Equations (\ref{Prad}) - (\ref{Ptang}) provide the expressions
for the radial and tangential pressures as
\begin{equation}
P_{r}=\frac{C}{8\,\pi}\frac{1-2x-K\left(  1+x+2x^{2}\right)  }{\left(
1+2x\right)  ^{2}}\qquad\mathrm{and}%
\end{equation}%
\begin{equation}
P_{\bot}=\frac{C}{8\pi}\frac{\left(  4x^{4}+6x^{3}+10x^{2}+7x-1\right)
K^{2}x-\left(  4x^{4}+24x^{3}+19x^{2}+4x+1\right)  K-6x^{2}-3x+1}{\left(
1+2x\right)  ^{3}\left(  1+x\right)  \left(  1+Kx\right)  }\ ,
\end{equation}
where $x=C\,r^{2}$ $;$the constants $K$ and $C$ are obtained from the boundary
conditions $M~=~m(a)$ and $P_{r}(a)~=~0$ respectively, i.e.
\begin{equation}
K=\frac{{x_{1}}^{\frac{3}{2}}-2MC^{\frac{1}{2}}\left(  1+2x_{1}\right)
}{x_{1}^{\frac{3}{2}}\left(  1+x_{1}\right)  }\qquad\mathrm{and}\qquad
M=\frac{1}{C^{\frac{1}{2}}}\frac{x_{1}^{\frac{5}{2}}}{1+x_{1}+2x_{1}^{2}}\ ,
\end{equation}
and $x_{1}=C\,a^{2}$

\paragraph{Example 5:}

The density profile of this example corresponds to a solution originally
proposed M. Wyman \cite{Wyman1949}. Again, the same solution is found in
\cite{Korkina1981,Durgapal1982,Kuchowicz1970,Adler1974} and \cite{AdamsCohen1975}%

\begin{equation}
\rho=-\frac{C}{8\,\pi}\frac{K\left(  3+5\,x\right)  }{\left(  1+3\,x\right)
^{\frac{5}{3}}}\qquad\Leftrightarrow\qquad m=-\frac{1}{2C^{\frac{1}{2}}}%
\frac{Kx^{\frac{3}{2}}}{\left(  1+3x\right)  ^{\frac{2}{3}}}\ , \label{Examp5}%
\end{equation}
with $x=C\,r^{2}$ and $K,C=const.$

The method applied to the density profile (\ref{Examp5}) leads to radial and
tangential pressures
\begin{equation}
P_{r}=\frac{C}{8\,\pi}\frac{K\left(  x-1\right)  }{\left(  1+3x\right)
^{\frac{5}{3}}}\qquad\mathrm{and}%
\end{equation}%
\begin{equation}
P_{\bot}=\frac{CK}{8\pi}\frac{\left(  1+3x\right)  ^{\frac{2}{3}}\left(
x^{2}+4x-1\right)  +Kx\left(  3x^{2}+8x+1\right)  }{\left(  1+3x\right)
^{\frac{8}{3}}\left[  Kx+\left(  1+3x\right)  ^{\frac{2}{3}}\right]  }\ ,
\end{equation}
where
\begin{equation}
K=-2^{\frac{7}{3}}\frac{M}{a}\qquad\mathrm{and}\qquad x_{1}=C\,a^{2}\ .
\end{equation}
The constant $K$ has been obtained for boundary conditions and the pressure at
the surface ($P_{r}(a)~=~0$) determines the next integration constant:
$C:=\frac{1}{a^{2}}$

\paragraph{Example 6:}

The following density profile was found by M.P. Korkina,\ in 1981,
\cite{Korkina1981} and rediscovered a year later by M.C. Durgapal
\cite{Durgapal1982}. It can be written as%

\begin{equation}
\rho=\frac{C}{8\,\pi\left(  1+\,x\right)  ^{2}}\left[  \frac{3}{2}\left(
3+x\right)  -\frac{3K\left(  1+3\,x\right)  }{\left(  1+4\,x\right)
^{\frac{3}{2}}}\right]  \Leftrightarrow m=-\frac{x^{\frac{3}{2}}}{4C^{\frac
{1}{2}}}\frac{2K-3\left(  1+4x\right)  ^{\frac{1}{2}}}{\left(  1+x\right)
\left(  1+4x\right)  ^{\frac{1}{2}}}; \label{Examp6}%
\end{equation}
with $x=C\,r^{2}$ and $K,C=const.$

Again, for the density profile (\ref{Examp6}) the radial and tangential
pressures are found to be
\begin{equation}
P_{r}=\frac{C}{8\,\pi}\frac{3\left(  1+4x\right)  ^{\frac{3}{2}}\left(
1-x\right)  -2K\left(  1-x-8x^{2}\right)  }{\left(  1+x\right)  ^{2}\left(
1+4x\right)  ^{\frac{3}{2}}}\qquad\mathrm{and}%
\end{equation}%
\begin{equation}
P_{\bot}=-\frac{C}{16\pi}\frac{4K^{2}x\,\Lambda+4K\left(  1+4x\right)
^{\frac{1}{2}}\,\Xi+3x\,\Pi+6}{\left(  1+x\right)  ^{3}\left(  1+4x\right)
^{\frac{5}{2}}\left[  \left(  1+4x\right)  ^{\frac{1}{2}}\left(  x-2\right)
-2Kx\right]  }\ ,
\end{equation}
where
\begin{align*}
\Lambda &  =-8x^{4}+34x^{3}+36x^{2}+13x+1\ ,\\
\Xi &  =8x^{5}-65x^{4}-16x^{3}+5x^{2}+x-1\qquad\mathrm{and}\\
\Pi &  =192x^{4}+80x^{3}+116x^{2}+87x+23\ .
\end{align*}

Finally, the constants $K$ and $C$ can be obtained for boundary conditions:
\begin{equation}
K=-\frac{\left(  1+4\,x_{1}\right)  ^{\frac{1}{2}}}{2\,x_{1}^{\frac{3}{2}}%
}\left[  4MC^{\frac{1}{2}}\left(  1+x_{1}\right)  -3\,x_{1}^{\frac{3}{2}%
}\right]  ;\qquad\mathrm{and}\qquad M=\frac{3\,x_{1}^{\frac{5}{2}}}%
{C^{\frac{1}{2}}\left(  8x_{1}^{2}+x_{1}-1\right)  }\ ,
\end{equation}
with $x_{1}=C\,a^{2}$

\subsection{Modeling Anisotropic Spheres with NLES}

The parameters: mass, $M$,in terms of solar mass $M_{\odot}$, $M/a$
gravitational potential at the surface, boundary redshift $z_{a}$, surface
density $\rho_{a}$ and central density $\rho_{c}$, that characterize these
bounded configurations are summarized in the following table

\begin{center}%
\begin{tabular}
[c]{|c|c|c|c|c|c|}\hline
\textbf{Equation of State} & $M/a$ & $M$ $(M_{\odot})$ & $z_{a}$ & $\rho_{a}$
$\times$ $10^{14}\,(gr.cm^{-3})$ & $\rho_{c}$ $\,\times10^{15}$ $(gr.cm^{-3}%
)$\\\hline
\textbf{Example 1} & 0.32 & 2.15 & 0.6 & 6.80 & 1.91\\\hline
\textbf{Example 2} & 0.40 & 2.80 & 1.2 & 8.84 & 1.99\\\hline
\textbf{Example 3} & 0.38 & 2.54 & 1.0 & 8.04 & 2.41\\\hline
\textbf{Example 4} & 0.25 & 1.69 & 0.4 & 5.36 & 2.00\\\hline
\textbf{Example 5} & 0.38 & 2.54 & 1.0 & 8.04 & 3.04\\\hline
\textbf{Example 6} & 0.35 & 2.37 & 0.8 & 7.75 & 2.11\\\hline
\end{tabular}
\end{center}

All these parameters have been tuned up in order to that the mass function
satisfies the physical and energy conditions, i.e., the inequalities
(\ref{mas2r}) thought (\ref{strgenerani1}) (or (\ref{mas2r}) thought
(\ref{velsonid}) and (\ref{domener}) and (\ref{domenetg})),\ for a typical
compact objects of radius of $a=10$ Km..

\section{$P_{r}=0,$ and $P_{\perp}=0$ static solutions:}

\label{magnetar}In this section we shall show that it is possible to obtain
credible anisotropic spherically matter configuration with a \textit{NLES} for
two particular cases: one where the radial pressure vanishes, i.e., $P_{\perp
}\neq0$ and $P_{r}=0$ and other where the tangential pressures vanishes, i.e.,
$P_{\perp}=0$ and $P_{r}\neq0.$

\subsection{$P_{r}=0$ and $P_{\perp}\neq0$}

The study of matter configurations with vanishing radial stresses traces back
to G. Lema\^{i}tre \cite{Lemaitre1933}, A. Einstein\cite{Einstein1939} and
P.S. Florides\cite{Florides1974} in the static case. Non static models have
been considered in the past in references
\cite{HerreraSantos1997,Datta1970,Bondi1971,Evans1977} and recently in
\cite{Magli1997,HaradaNakaoIguchi1999,JhiganMagli2000,BarveSinghWitten2000}
concerning their relation with naked singularities. More recently conformally
flat models with vanishing radial pressures has been considered in
\cite{HerreraEtal2001}. We shall present here one of these static solution
satisfying an \textit{NLES.}

From equation (\ref{Prad}) it is trivial to see that
\begin{equation}
\frac{2m^{\prime}}{r^{2}}=\frac{4m}{r^{3}}\qquad\Rightarrow\quad m\left(
r\right)  =Cr^{2}. \label{mPr0}%
\end{equation}
Thus, density and tangential pressure profiles can be calculated as%

\begin{equation}
\rho=\frac{C}{2\pi r}\qquad\mathrm{and}\qquad P_{\perp}=\frac{C^{2}}%
{4\pi(1-2Cr)}\ .
\end{equation}
The constant $C$ can be obtained from the boundary conditions, i.e.%

\begin{equation}
m(a)=M\qquad\mathrm{\Rightarrow}\qquad C=\frac{M}{a^{2}}\ .
\end{equation}

\subsection{$P_{\perp}=0$ and $P_{r}\neq0$}

The second static\ solution with a \textit{NLES} we shall consider is the case
$P_{\perp}=0$ and $P_{r}\neq0.$ The rationale behind this supposition is the
considerable effort that has been directed in recent years to study the
effects of intense magnetic fields ($B\gtrsim10^{15}$ G) on highly compact
astrophysical objects (\cite{SaengMathews1999}-\cite{KohriYamadaNagataki2001}
and references therein). In fact, some observations seems to confirm that
newly born neutron stars with very large surface magnetic fields (magnetar)
could represent either soft gamma repeaters or anomalous X Ray pulsars
\cite{SaengMathews2000}, or both. One of the most striking effects of these
large magnetic fields on the equation of state for superdense matter
distributions is that they induce local anisotropic
pressures\cite{ChaichianEtal1999,PerezMartinezEtal2000,KohriYamadaNagataki2001}
and it is possible to obtain extreme cases were the pressure perpendicular to
the magnetic field could vanish. In our case we shall suppose that tangential
stresses vanish in all directions. In this sense we consider our solution
inspired by the magnetar models

It is clear that if $P_{\perp}=0$ equation (\ref{Ptang}) leads to
\begin{equation}
\frac{m^{\prime\prime}}{r}+\frac{2\left(  m^{\prime}r-m\right)  }{r^{3}%
}\left[  \frac{m^{\prime}r-m}{r-2m}-1\right]  =0\ , \label{edemesinpt}%
\end{equation}
which can be integrated yielding%

\begin{equation}
m=\frac{r}{2}\left[  1-e^{-2\left(  C_{1}r+C_{2}\right)  }\right]  \ .
\label{emesinpt}%
\end{equation}
The density and the pressure profiles emerge from equations (\ref{eme}) and
(\ref{globalst}), respectively as:%

\begin{align}
\rho &  =\frac{e^{-2(C_{1}r+C_{2})}}{8\pi r^{2}}\left[  2rC_{1}-1+e^{2(C_{1}%
r+C_{2})}\right]  \qquad\text{and}\\
P_{r}  &  =\frac{e^{-2(C_{1}r+C_{2})}}{8\pi r^{2}}\left[  2rC_{1}%
+1-e^{2(C_{1}r+C_{2})}\right]  \ .
\end{align}
The constants $C_{1}$ and $C_{2}$ can be obtained for boundary conditions at
$r=a$ as%
\begin{align}
m(a)  &  =M\qquad\mathrm{\Rightarrow}\qquad C_{1}=\frac{M}{a\left(
a-2M\right)  }\qquad\text{and}\\
P_{r}(a)  &  =0\qquad\mathrm{\Rightarrow}\qquad C_{2}=\frac{-M}{a-2M}+\frac
{1}{2}\ln\left(  \frac{a}{a-2M}\right)  \ .
\end{align}

It is worth mentioning that from equation (\ref{emesinpt}), we have

\begin{equation}
m=0\qquad\mathrm{\Rightarrow}\qquad\left\{
\begin{array}
[c]{l}
r=0\\
\\
r=\frac{a}{2}\left[2+\left(\frac{a}{M}-2\right)\ln\left(1-2\frac{M}{a}\right)\right]
\end{array}
\right.  \label{cerosdeme}%
\end{equation}
Therefore. matter configurations with \textit{NLES} having vanishing
tangential pressures are only possible within a region of the sphere, i.e.
\begin{equation}
\frac{a}{2}\left[2+\left(\frac{a}{M}-2\right)\ln\left(1-2\frac{M}{a}\right)\right]  ~<~r~<~a,
\end{equation}
because$,$within the range $0~<~r~<~\frac{a}{2}\left[2+\left(\frac{a}{M}
-2\right)\ln\left(1-2\frac{M}{a}\right)\right]  $ the mass function becomes negative.

\section{Concluding Remarks}

\label{concluding}We presented a method to obtain \textit{NLES} static
anisotropic spherically symmetric exact solutions starting from known density
profiles. It is clear that, when such a density profile, $\rho(r)$, is
provided the radial pressure, $P_{r}(r)$, can be obtained from the
\textit{NLES}(\ref{globalst}) and the tangential pressure $P_{\perp}(r)$ can
be solved algebraically from the anisotropic Tolman-Oppenheimer-Volkov (TOV)
differential equation (\ref{anitov}). Althought, we supposed to have found a
more general family of solutions for the Einstein Equations, (\ref{beta}),
than the one considered in \cite{HernandezNunezPercoco1998} but, due to the
boundary conditions (\ref{zeroc}), the actual family of solutions is indeed
the previous (\ref{oldsoluc}) considered in reference (\ref{oldsoluc}).

It is also obtained credible anisotropic spherically matter configurations
with a \textit{NLES} for two particular cases: one where the radial pressure
vanishes, i.e., $P_{\perp}\neq0$ and $P_{r}=0$ and other where the tangential
pressures vanishes, i.e., $P_{\perp}=0$ and $P_{r}\neq0.$

We worked out in detail the junction and energy conditions for anisotropic
(and isotropic) bounded matter distributions, i.e. (\ref{mas2r}) thought
(\ref{domenetg}), establishing their consequences in terms of the mass
function $m(r)$ for bounded configurations having \textit{NLES.} In the early
days of Relativistic Astrophysics, some of these results led to the first
general theorems due to H. A. Buchdahl\cite{Buchdahl1959,Buchdahl1981} and H.
Bondi\cite{Bondi1964,TrautmanPiraniBondi1965} concerning inequalities limiting
the behavior of the mass function for compact objects.

Because the mass function $m(r),$ presented in equations (\ref{Examp1})
through (\ref{Examp6}) and the corresponding pressure profiles are regular at
the origin, the solutions to the Einstein Equations are considered unique in
the sense stated by Rendall and
collaborators\cite{RendallSchmidt91,BaumgarteRendall94}. That is to say: for a
given value of the central pressure, $P_{0}=P_{\perp}(r=0)=P_{r}(r=0)$, there
exists a unique global solution $P_{r}(r)$ to the anisotropic TOV equation
(\ref{anitov}) representing a spherically symmetric fluid sphere in General Relativity.

It is clear that the last term in the anisotropic TOV equation (\ref{anitov}),
$\frac{2}{r}\left(  P_{\perp}-P_{r}\right)  \equiv\Delta,$ represents a
``force'' due to the local anisotropy. This ``force'' is directed outward when
$P_{\perp}>P_{r}\Leftrightarrow\Delta>0$ and inward if $P_{\perp}%
<P_{r}\Leftrightarrow\Delta<0$. As it is apparent from the figure and also
from the above table, those models with a (average) repulsive ``anisotropic''
force ($\Delta>0$) allows the construction of more massive distributions. This
picture is more evident for solutions where $P_{\perp}\neq0$ and $P_{r}=0$ and
$P_{\perp}=0$ and $P_{r}\neq0,$ which are the extreme situations. Considering
plausible static matter configurations (it is to say those that satisfy
conditions (\ref{mas2r}) thought (\ref{domenetg})) we have

\begin{itemize}
\item  more massive configurations if the radial pressure vanishes i.e.
$\Delta>0$ has a maximum,

\item  less massive configurations if the tangential pressures vanishes i.e.
$\Delta<0$ has a minimum.
\end{itemize}

Notice that Model 4 (\textit{NLES }Tolman IV anisotropic static solution)\ and
Model 5 (\textit{NLES} anisotropic Wyman like static solution) have ``soft''
($P_{\perp}<P_{r}$) cores and ``hard'' ($P_{\perp}>P_{r}$) outer mantles. It
is worth mentioning that models presented here are very sensitive to the
change of values of the parameters sketched in the table.

As it can be appreciated from equation (\ref{cerosdeme}), fluids satisfying
\textit{NLES} and having vanishing tangential pressures are only possible
within a region of the matter configuration with $\frac{a}{2}\left[
2+\left(\frac{a}{M}-2\right)\ln\left(1-2\frac{M}{a}\right)\right]  ~<~r~<~a.$
Within the inner core,
the mass function becomes negative. This situation emerges from a limitation
of the definition of the Schwarzschild mass function (\ref{eme}) when it is
apply to anisotropic fluid configurations. It is interesting to say that the
expression of the Tolman Whittaker mass\cite{Tolman1930}%
\begin{equation}
m_{TW}\left( r\right)  =4\pi\int_{0}^{r}r^{2}{\large e}^{\left(  \nu
+\lambda\right)  /2}\left(  \rho-P_{r}-2P_{\perp}\right)  \mathrm{d}r\equiv
e^{\left(  \nu+\lambda\right)  /2}\left(  m+4\pi P_{r}r^{3}\right)
\label{massTW}%
\end{equation}
for this static \textit{NLES} model reads
\begin{equation}
m_{TW}\left(  r\right)  =C_{1}\ r^{2}\label{mTWPt0}%
\end{equation}
and precludes non physical situations at the core of the distribution. It is
clear that for this case the Tolman-Whittaker mass function, $m_{TW},$ is a
more suitable concept of mass than the Schwarzschild mass function, $m$ (see
\cite{HerreraSantos1997} for a comprehensive discussion on this point,
plentiful of examples). More over, from equations (\ref{mPr0}) and
(\ref{mTWPt0}), the expression of the $m_{TW}$ for this particular static
\textit{NLES} model with vanishing tangential stresses is the same that $m$
for the model with vanishing radial pressure.

It can also be mentioned as a curiosity that the relation among the areas
under the curves displayed in the figure, i.e. $\int_{0}^{R}\frac
{2\mathrm{d}r}{r}\left(  P_{\perp}-P_{r}\right)  ,\ $corresponds to the
relation among de total masses. The more massive the model is the greater the
area under the curve $\frac{2}{r}\left(  P_{\perp}-P_{r}\right)  $ is.

In principle, a \textit{NLES} in isotropic pascalian fluid ($P_{\perp}=P_{r}$)
can not be ruled out, but we have integrated numerically the standard TOV
equation (\ref{anitov}), which can be written in terms of the mass
distribution function as
\begin{equation}
m^{\prime\prime}=\frac{2m^{\prime}}{r}-\frac{4m}{r^{2}}-\frac{2\left(
m^{\prime}r-m\right)  }{r^{2}}\left[  \frac{m^{\prime}r-m}{r-2m}-1\right]
\qquad\mathrm{with}\qquad\left\{
\begin{array}
[c]{c}%
m(0)=0\\
\\
m^{\prime}(0)=0
\end{array}
\right.
\end{equation}
and the inequalities (\ref{mas2r}) thought (\ref{strgenerani1}) (or
(\ref{mas2r}) thought (\ref{velsonid}) and (\ref{domener}) \& (\ref{domenetg}%
)) corresponding the physical and energy conditions \textbf{were not}
fulfilled for any configuration having a \textit{NLES}.

\section{Acknowledgments}

We are indebted to J. Fl\'{o}rez L\'{o}pez for pointing out us the relevance
of nonlocal theories in modern classical continuum mechanics and to L. Herrera
Cometta for the references of Magnetars. We also gratefully acknowledge the
financial support of the Consejo de Desarrollo Cient\'{i}fico Human\'{i}stico
y Tecnol\'{o}gico de la Universidad de Los Andes under project C-1009-00-05-A,
and to the Consejo Nacional de Investigaciones Cient\'{i}ficas y
Tecnol\'{o}gicas under project S1-2000000820

 \section*{Figure caption}

\noindent Figure 1: $\frac{2}{r}\left(  P_{\perp}-P_{r}\right)\equiv\Delta
,$ as function of $r/a$ which represents a ``force'' due to the local
anisotropy. This ``force'' is directed outward when $P_{\perp}>P_{r}%
\Leftrightarrow\Delta>0$ and inward if $P_{\perp}<P_{r}\Leftrightarrow
\Delta<0$. Model 4 (\textit{NLES }Tolman IV anisotropic static solution)\ and
Model 5 (\textit{NLES} anisotropic Wyman like static solution) have ``soft''
($P_{\perp}<P_{r}$) cores and ``hard'' ($P_{\perp}>P_{r}$) outer mantles.

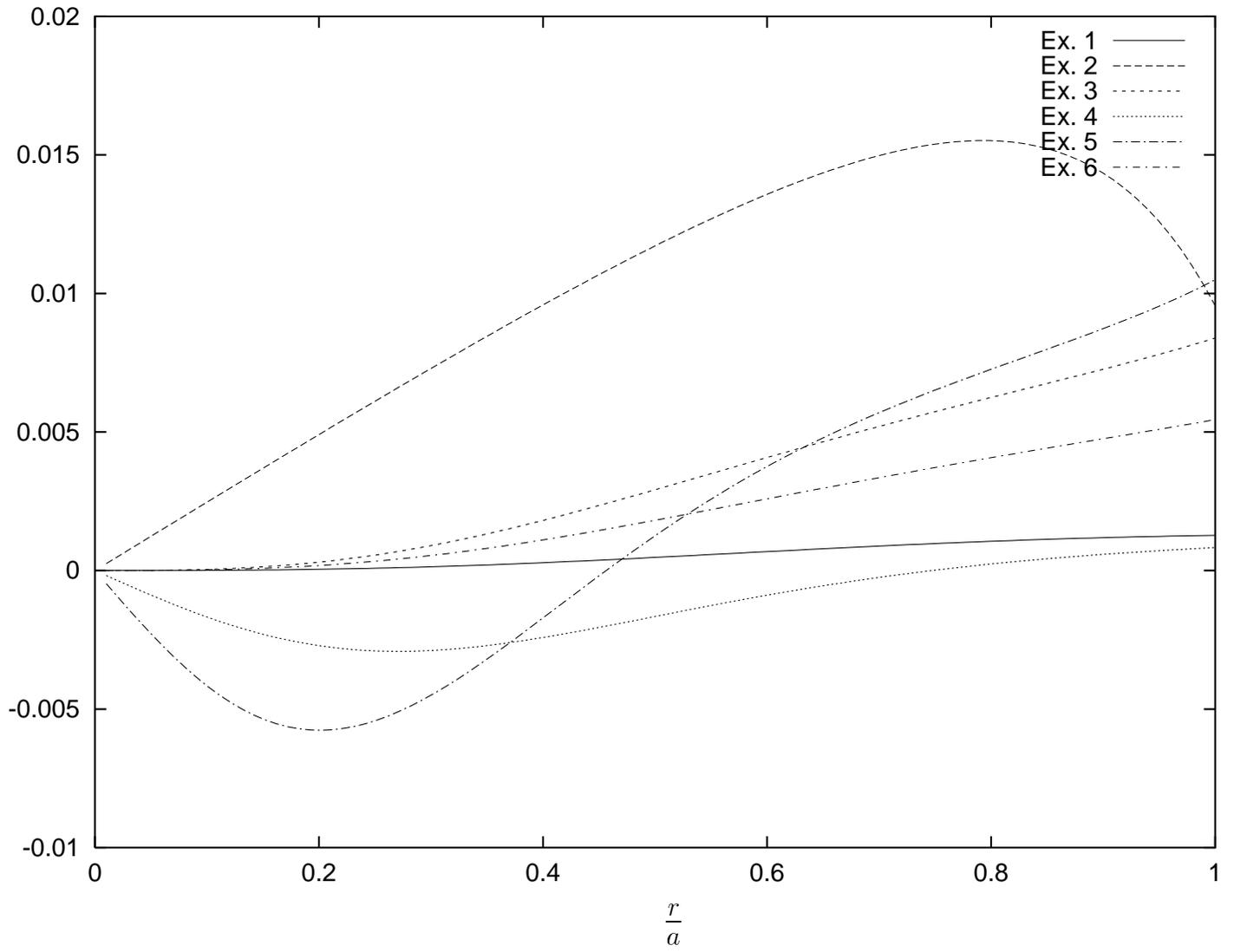
\end{document}